\documentstyle[12pt,epsf]{article}
\normalbaselines

\newcommand{\be}{\begin{equation}}
\newcommand{\ee}{\end{equation}}
\newcommand{\bea}{\begin{eqnarray}}
\newcommand{\eea}{\end{eqnarray}}

\title{Exactly solvable hydrogen-like potentials and factorization method}

\bigskip \bigskip

\author{J. Oscar Rosas-Ortiz\thanks{On leave of absence from 
         {\it Departamento de F\'\i sica}, CINVESTAV-IPN, 
         {\it A.P. 14-740, 07000 M\'exico D.F., Mexico}. 
           E-mail orosas@fis.cinvestav.mx}\\
       \small {\it Departamento de F\'\i sica Te\'orica\thanks{Electronic 
           address: orosas@klander.fam.cie.uva.es}}\\
       \small {\it Universidad de Valladolid, 47011 Valladolid, 
                                    Spain}}

\bigskip \bigskip

\date{}

\begin{document}

\maketitle

\thispagestyle{empty}
\begin{abstract}

A set of factorization energies is introduced, giving rise to a
generalization of the Schr\"{o}dinger (or Infeld and Hull) factorization
for the radial hydrogen-like Hamiltonian. An algebraic intertwining
technique involving such factorization energies leads to derive
$n$-parametric families of potentials in general almost-isospectral to the
hydrogen-like radial Hamiltonians. The construction of SUSY partner
Hamiltonians with ground state energies greater than the corresponding
ground state energy of the initial Hamiltonian is also explicitly
performed. 
\end{abstract}

\bigskip\bigskip

{\footnotesize
\begin{description}
\item[Key-Words:] Factorization, Hydrogen atom
\item[PACS:] 03.65.Ge, 03.65.Fd, 03.65.Ca 

\end{description}}

\vfill


\newpage

\setcounter{page}{1}

\newpage 

\section{Introduction} 

The factorization method has been historically atributed to
Schr\"{o}dinger \cite{S40}. This approach (originated from the well known
treatment of the harmonic oscillator in non-relativistic Quantum
Mechanics) avoids the use of cumbersome mathematical tools. Since
Schr\"{o}dinger the factorization method has been successfully applied to
solve essentially any problem for which there exists exact solution
\cite{IH51}. Considering the narrow set of exactly solvable potentials, it
is interesting to look for new potentials for which the corresponding
Schr\"{o}dinger equation becomes analitically solvable. There are a lot of
results on the matter, in particular it is worthwhile mentionning a work
of Mielnik \cite{M84}, where a variant of the standard factorization was
introduced in order to get a family of Hamiltonians isospectral to the
harmonic oscillator. The first application of Mielnik's factorization was
performed for the hydrogen-like radial potentials \cite{F84}, the
resulting family of Hamiltonians became isospectral to the standard radial
hydrogen-like Hamiltonians. An additional interesting application arose
from the arena of the coherent states. Concerning this subject, it is well
known that the usual first order factorization operators for the harmonic
oscillator Hamiltonian are also its creation and anihilation operators. 
This is not the case for the isospectral oscillator Hamiltonians derived
by Mielnik because their first order factorization operators are different
from their creation and anihilation operators, which turn out to be of
order equal to (or greater than) three \cite{M84,F94}. These last
operators are the generators of non-linear algebras, and allow one to get
immediately the corresponding coherent states as the eigenstates of the
annihilation operator \cite{F94}. Recently, these algebras have been
rederived (using different methods) and reinterpreted by other authors
\cite{SBL98,JR97}.  It has been shown as well that such non-linear
algebras can be associated to other systems \cite{JR97}. 

The factorization techniques \cite{S40,IH51,M84}, the Darboux
transformation \cite{D882}, the supersymmetric Quantum Mechanics (SUSY QM)
\cite{W81}, and other procedures have a common point: they can be embraced
in the algebraic scheme known as {\it first order intertwining
transformation} which makes use of a first order differential operator to
{\it intertwine} (the eigenfunctions of)  two different Hamiltonians
\cite{C79}. The generalization of this technique involves higher order
differential operators and leads in a natural way to the higher order SUSY
QM [11-15].  One of the advantages of the intertwining technique is that
the composition of $n$ first order intertwining transformations allows one
to express the $nth$-order intertwiner operators as the product of $n$
first order intertwining operators \cite{AB85,R97,F98}. 

The experimental study of spectra (for systems as nuclei, atoms,
molecules, etc) makes important {\it per se} the theoretical determination
of the energy levels of potentials capable of being used as models for
describing different physical situations. However, for a quantum system
with a given Hamiltonian it is seldom possible to get exact solutions to
the corresponding eigenproblem. 

In this paper we are going to show that, given an exactly solvable
potential as an input, the {\it intertwining technique} provides, in
general, new exactly solvable potentials which, from the traditional
viewpoint, would be solved by perturbative or approximate methods. Hence,
working with Hamiltonians whose spectrum is known, one has at hand the
tools to test the validity and convergence of some perturbative or
approximate methods \cite{R97,note1}. In Section 2 we shall present the
$nth$-order intertwining technique, discussing in more detail those
aspects which have not been previously reported in the literature. In
particular, in Section 2.1 we shall show that the standard Infeld-Hull
factorization as well as the modified Mielnik's method can be recovered by
the first order intertwining treatment. We will introduce a generalization
of the Mielnik's approach by using solutions to the Schr\"{o}dinger
equation corresponding to factorization energies not belonging to the
physical spectrum of the initial problem. In Section 2.2, by using the
second order intertwining approach, we will introduce explicit expressions
allowing the construction of {\it two}-parametric families of potentials
whose spectrum is equal to the initial spectrum plus two new levels at
predetermined positions. In Section 2.3 we shall prove that the second
order, as well as the higher order cases, can be recovered by the
iterative application of our first order intertwining procedures. Section
3 contains the derivation of a wide set of new $n$-parametric families of
potentials whose spectrum is the same (isospectral), or almost the same
(almost isospectral), as the corresponding hydrogen-like spectrum. We
shall introduce a set of factorization energies generalizing the choice
made for the Infeld-Hull and Mielnik factorizations of the hydrogen-like
potentials. As a consequence, the first $n$ energy levels of the
$n$-parametric families can be placed at any previously fixed spectral
positions chosen from the above mentionated set of factorization energies. 
This result opens the possibility of placing {\it holes} between these
levels or between them and the other ones. Some particular cases related
to results previously derived by other authors are also mentioned.

\section{The Intertwining Method}

Let $H$ and $\widetilde H$ be the one dimensional Hamiltonians
\be
H \equiv -\frac{d^2}{dx^2} + V(x), \quad \quad \widetilde H \equiv
-\frac{d^2}{dx^2} + \widetilde V(x).
\label{1}
\ee
The {\it nth}-order {\it intertwining method} aims to find a {\it
nth}-order differential operator $A$ such that
\be
\widetilde H A = A H.
\label{2}
\ee

Let $H$ be a Hamiltonian with known solutions to the time-independent
Schr\"{o}dinger equation $H \psi = E \psi$. We are looking for solutions
to the corresponding equation for the {\it intertwined Hamiltonian}
$\widetilde H$, $\widetilde H \widetilde \psi = \widetilde E \widetilde
\psi$. The transformation (\ref{2}) preserves many spectral properties of
the initial Hamiltonian, {\it e.g.} $\widetilde H ( A \psi)  = E (A
\psi)$, $\forall \ A \psi \neq 0$. We are interested only in those
operators $A$ whose action on $L^2({\bf R})$ produces functions in
$L^2({\bf R})$; hence we can choose $\widetilde \psi \equiv c_0 \, A
\psi$, with $c_0$ a normalization constant, and we will have $\widetilde E
= E, \ \forall \ \widetilde \psi =c_0 A \psi \neq 0 , \psi \in L^2 ({\bf
R})$. Therefore, the intertwining operator $A$ transforms solutions $\psi$
of the initial Schr\"{o}dinger equation into solutions $\widetilde \psi$
of a new Schr\"{o}dinger equation, both equations sharing the initial
eigenvalues. However, the new Hamiltonian could have a finite number of
additional eigenvalues; if this is the case, the corresponding
eigenfunctions have to be elements of the kernel of $A^\dagger$, as we
shall see below. We will show now how the intertwining method does work
for the simplest case where $A$ is a first order differential operator.
This case is specially interesting because it provides a general framework
in which various apparently different methods are included. 

\subsection{The first order intertwining method}

Let the operator involved in the intertwining relationship be the
following first order differential operator
\be
A \equiv a = \frac{d}{dx} + \beta(x).
\label{3}
\ee
By introducing (\ref{1}) and (\ref{3}) in (\ref{2}) we get
\be
- \beta'(x) + \beta^2(x) = V(x) - \epsilon,
\label{4}
\ee
\be
\widetilde V(x) = V(x) + 2 \beta'(x),
\label{5}
\ee
where $\epsilon$ is an arbitrary integration constant, and the prime
denotes derivative with respect to $x$. Notice the implicit dependence of
$\beta(x)$ and $\widetilde V (x)$ on $\epsilon$.

From equations (\ref{1}) and (\ref{5}) it becomes apparent that
$\widetilde H -H = 2 \beta'(x)$, from (\ref{3}) we get $[a, a^\dagger] = 2
\beta'(x)$, therefore (\ref{4})  and (\ref{5}) lead to: 
\be
H= a^\dagger a + \epsilon, \quad \quad \widetilde H = a a^\dagger +
\epsilon.
\label{6}
\ee
Equation (\ref{6}) shows that the first order intertwining relationship
(\ref{2}), with $A$ given by (\ref{3}), leads in a natural way to the
factorization of the intertwined Hamiltonians for each value of $\epsilon
\in {\bf R}$, provided that $\beta(x)$ satisfies (\ref{4}) with a given
$V(x)$. The traditional approach to the factorization method involves just
a particular solution to the Riccati differential equation (\ref{4}) for a
specific value of $\epsilon$ \cite{IH51}. However, it is well known that a
particular solution to a Riccati equation allows one to get the
corresponding general solution by means of two quadratures \cite{I26}. 
Mielnik's approach takes this into account \cite{R97} and gives a term in
$\beta(x)$ unnoticed by the traditional factorization method. With the
exception of the harmonic oscillator and the hydrogen-like potentials, the
general solutions to the corresponding Riccati equation (\ref{4}) remain
unexplored for almost all the Infeld and Hull cases \cite{note2,note2a}.
An interesting discussion on the relationship between Mielnik's method and
Darboux transformation was given by Andrianov {\it et al} in 1985
\cite{AB85} (see also \cite{note2a}). On the other hand, equations
(\ref{4}) and (\ref{5}) are basic in the construction of the SUSY partner
Hamiltonians $H^+ \equiv H - \epsilon = a^\dagger a$, and $H^- \equiv
\widetilde H - \epsilon = a a^\dagger$ in supersymmetric quantum
mechanics. In this language, it is said that equation (\ref{5}) relates
the two SUSY partner potentials $\widetilde V(x)$ and $V(x)$, and the
function $\beta(x)$ is named the SUSY potential \cite{W81}. Notice that
the two Hamiltonians $H^\pm$ are positive semi-definite, {\it i.e.}
$\langle H^\pm \rangle \geq 0$; hence we have $E \geq \epsilon$ and
$\widetilde E \geq \epsilon$.

If the eigenfunctions $\{\psi\}$ and eigenvalues $\{E\}$ of $H$ are known,
the corresponding eigenfunctions of $\widetilde H$ are given by
$\widetilde \psi = (E - \epsilon)^{-1/2} \, a \psi$ with eigenvalues
$\widetilde E = E$. They form an orthonormal set 
\[
\langle \widetilde \psi', \widetilde \psi \rangle =
[(E'-\epsilon)(E-\epsilon)]^{-1/2} \langle a \psi', a \psi \rangle =
\sqrt{\frac{E'-\epsilon}{E-\epsilon}} \langle \psi', \psi\rangle = 0,
\quad \forall \psi' \neq \psi.
\]
However, they do not span yet the whole of $L^2 ({\bf R})$ because the
solution to $a^\dagger \widetilde \psi_{\epsilon}= 0$ is orthogonal to all
the members of the set $\{\widetilde \psi\}$, $ \langle \widetilde
\psi_{\epsilon}, \widetilde \psi \rangle = (E-\epsilon)^{-1/2}\langle
a^\dagger \widetilde \psi_{\epsilon}, \psi \rangle = 0$. That first order
linear differential equation is immediately solved: 
\be
\widetilde \psi_{\epsilon} \propto e^{\int \beta(x) dx}.
\label{7}
\ee
From (\ref{6}), it is easy to see that $\widetilde H \widetilde
\psi_{\epsilon} = \epsilon \widetilde \psi_{\epsilon}$. Hence, $\widetilde
\psi_{\epsilon}$ is an eigenfunction of $\widetilde H$ with eigenvalue
$\epsilon$. In the case when $\widetilde \psi_{\epsilon}$ is a square
integrable function ({\it i.e.} $\widetilde \psi_{\epsilon}$ has physical
sense as a wavefunction), it has to be added to the set $\{\widetilde
\psi=(E-\epsilon)^{-1/2} a \psi\}$ in order to complete a new basis on
$L^2 ({\bf R})$. The set of eigenvalues of $\widetilde H$ is given by
$\{\widetilde E\} = \{\epsilon, E\}$, so that the Hamiltonian $\widetilde
H$ has the same spectrum as $H$ plus a new level at $\epsilon$. Indeed,
since the general solution $\beta(x)$ of the Riccati equation (\ref{4})
has a `{\it free parameter}' $\lambda$, labeling also $\widetilde V(x)$
(see equation (\ref{5})), the Mielnik's method allows one to construct,
for a specific value of $\epsilon$, a {\it one}-parametric family of
potentials $\widetilde V(x)$ which in general will be almost isospectral
to $V(x)$. 

We are going to disscus now how the Mielnik's method can be further
generalized. First, let us remark that the factorization energy considered
by Infeld-Hull and Mielnik is not the only posibility leading to a
solvable Riccati equation (\ref{4}). The key point arises when the
non-linear first order differential equation (\ref{4}) is replaced into a
homogeneous linear second order differential equation by means of the
transformation\footnote{The inverse transformation of (\ref{8}) is
$u(x)=\exp(-\int\beta(x)dx)$.}:
\be
\beta(x) = - \frac{d}{dx} \ln u(x),
\label{8}
\ee
which leads to the second order differential equation:
\be
\frac{d^2}{dx^2}u(x) - [V(x) - \epsilon] u(x) =0.
\label{9}
\ee
If the factorization energy $\epsilon$ belongs to $\{ E \}$, then the
solutions $u(x)$ to (\ref{9}) are the same as the previously known
eigenfunctions $\psi(x)$ of the initial Hamiltonian $H$.  Following
\cite{M84}, we shall consider just the cases where the factorization
energy $\epsilon$ does not belong to the spectrum of $H$. Therefore, the
solutions to (\ref{9})  do not have direct physical meaning, but we have
just seen that they naturally lead to the factorization of the
Hamiltonians $H$ and $\widetilde H$ in the spirit of Mielnik's approach,
providing as well the explicit form for the new potentials $\widetilde
V(x)$. 

Let us remark that $n$ different factorization energies $\epsilon_1,
\epsilon_2, ...,\epsilon_n$, lead to $n$ different Riccati equations
(\ref{4}) and to $n$ {\it one}-parametric families of potentials
$\widetilde V_i(x)$, $i=1,2,...,n$, almost isospectral to $V(x)$. It must
be clear that there will be $n$ different factorizations of the initial
Hamiltonian $H$ (see equation (\ref{6})). With the aim of iterating later
the first order intertwining technique let us introduce a slightly
different notation. Let us write $\epsilon = \epsilon^{(k)}$, where $k$
labels the specific real number $\epsilon \not \in \{ E \}$, which we will
use as factorization energy. Moreover, we shall write explicitly the
dependence of $\beta(x)$ and $\widetilde V(x)$ on $\epsilon^{(k)}$ (see
equations (\ref{4}) and (\ref{5})) by making $\beta(x) = \beta^{(k)}(x)$
and $\widetilde V(x) = V^{(k)}(x)$. For the sake of simplicity the other
functions and operators labeled by a {\it tilde} will be represented as
$\widetilde H = H^{(k)}$, $\widetilde E = E^{(k)}$, etc. Finally, the
number of indices will denote the order of the intertwining transformation
(\ref{2}), as well as the number of `{\it free parameters}' $\lambda$
which will label the intertwined potential. For example, $H^{(k)}$
represents a first order intertwined Hamiltonian whose corresponding {\it
one}-parametric family of potentials $V^{(k)}(x)$ is almost isospectral to
$V(x)$, while $H^{(km)}$ represents a second order intertwined Hamiltonian
with the corresponding {\it two}-parametric family of potentials $
V^{(km)}(x)$ almost isospectral to $V(x)$. The benefits of such notation
will be clear in the next sections.


\subsection{Second order intertwining method}

Let us consider the second order differential operator
\be
B^{(km)} = \frac{d^2}{dx^2} + \eta (x) \frac{d}{dx} +
\gamma(x), 
\label{10}
\ee
where $\eta(x)$ and $\gamma(x)$ depend implicitly on $\epsilon^{(k)}$ and
$\epsilon^{(m)}$. Introducing $H^{(km)}= -d^2/dx^2+V^{(km)}(x)$ and
(\ref{10}) in the second order intertwining relationship
$H^{(km)}B^{(km)}= B^{(km)}H$, with $H$ given in (\ref{1}), and by
ordering at different powers of the operator $d/dx$, we have for the
corresponding coefficients
\be
V^{(km)}(x) = V(x) + 2 \eta'(x),
\label{12}
\ee
\be
2 \gamma(x) = \eta^2(x) - \eta'(x) - d -2 V(x),
\label{13}
\ee
\be
V''(x) + \eta(x) V'(x) = 2 \gamma(x) \eta'(x) - \gamma''(x),
\label{11a}
\ee
where $d$ is an integration constant. Following \cite{F97}, we can express
the functions $V(x)$, $V^{(km)}(x)$ and $\gamma(x)$ in terms of $\eta(x)$.
The substitution of (\ref{13}) into (\ref{11a}) produces a third order
differential equation which, after multiplying by $\eta(x)$, becomes
reduced (by a first integration) to the second order differential equation
\be
\eta \eta'' -\frac{(\eta')^2}{2} + \left( 2 \gamma - \eta' -
\frac{\eta^2}{2} \right) \eta^2 + 2c=0,
\label{11}
\ee
where we have used once again the equation (\ref{13}) and $c$ is a new
integration constant.

In order to solve the non-linear second order differential equation
(\ref{11}), let us try the ansatz \cite{note3}
\be
\eta'(x) = \eta^2(x) + z(x) \eta(x) + b,
\label{14}
\ee
where the constant $b$ and the function $z(x)$ are to be determined. 
Introducing (\ref{14}) in (\ref{11}) we get $b_{\pm}= \pm 2 \sqrt{c}$,
$c>0$, besides two Riccati differential equations
\be
z_{\pm}'(x) +\frac{z_{\pm}^2(x)}{2} - 2V(x) -d - b_{\pm} = 0.
\label{15}
\ee
The identification $z_{+}(x) =- 2 \, \beta_k(x)$, $z_{-}(x) = -2 \,
\beta_m(x)$, and
\be
d= -\epsilon^{(m)} - \epsilon^{(k)}, \quad \quad c = \left(
\frac{\epsilon^{(m)} - \epsilon^{(k)}}{2} \right)^2, \quad \quad
\epsilon^{(m)}\neq \epsilon^{(k)},
\label{16}
\ee
allows one to write the solutions of (\ref{11}) in the form
\be
\eta(x) = - \left( \frac{\epsilon^{(m)} - \epsilon^{(k)}}{\beta^{(m)}(x) 
- \beta^{(k)}(x)} \right), \quad \quad \epsilon^{(m)} \neq \epsilon^{(k)},
\label{17}
\ee
where $\beta^{(m)}(x)$ and $\beta^{(k)}(x)$ are both solutions to
(\ref{4}) with factorization energies $\epsilon^{(m)}$ and
$\epsilon^{(k)}$ respectively. Equations (\ref{17}) and (\ref{13}) 
determine the second order intertwining operator $B^{(km)}$. Notice that,
by means of this process, we can derive solutions to the non-linear second
order differential equation (\ref{11}) by just solving the easier
non-linear first order differential equation (\ref{4}) for {\it two
different} values of the factorization energy $\epsilon$. 

The action of $B^{(km)}$ on the eigenfunctions $\{ \psi\}$ of $H$ gives
the basic set of eigenfunctions $\{ \Psi^{(km)} \propto B^{(km)} \psi\}$
of the intertwined Hamiltonian $H^{(km)}$. On the other hand, the
solutions to the second order differential equation ${B^{(km)}}^\dagger
\Psi^{(km)}_B=0$ are orthogonal to all the $\Psi^{(km)} \propto B^{(km)}
\psi$. Now, because the kernel of ${B^{(km)}}^\dagger$ is a 2-dimensional
subspace, the maximum number of linearly independent elements is two.
Therefore, we are looking for two square integrable linearly independent
solutions to ${B^{(km)}}^\dagger \Psi^{(km)}_B=0$ such that they are
simultaneously eigenfunctions of $H^{(km)}$, with eigenvalues to be
determined.

Let us now write $\Psi^{(km)}_B= {\cal C}_0 \exp( \int f^{(km)}(x) dx)$ as
a generic kernel element of ${B^{(km)}}^\dagger$; then the equation
${B^{(km)}}^{\dagger} \Psi^{(km)}_B = 0$ can be rewritten as
\be
\frac{d}{dx}f^{(km)}(x) -\eta(x) f^{(km)}(x) + (f^{(km)}(x))^2 -
\eta'(x) +
\gamma(x) =0. 
\label{18}
\ee
Using equations (\ref{4}), (\ref{13}) and (\ref{14}), the general
solutions to (\ref{18}) can be obtained:
\be
\Psi^{(km)}_B(x) = {\cal C}_k \frac{u^{(k)}(x)}{W(k,m)}+ {\cal C}_m
\frac{u^{(m)}(x)}{W(k,m)},
\label{19}
\ee
where $u^{(k)}(x)$ and $u^{(m)}(x)$ are the two {\it unphysical} solutions
to (\ref{9}) with eigenvalues $\epsilon^{(k)}$ and $\epsilon^{(m)}$,
respectively.  The function $W(k,m) = u^{(k)} (u^{(m)})' -(u^{(k)})'
u^{(m)}$ is the wronskian of $u^{(k)}(x)$ and $u^{(m)}(x)$. 

By using (\ref{8}) and (\ref{9}) we now rewrite (\ref{17}) as
\be
\eta(x) = ( \epsilon^{(m)} - \epsilon^{(k)}) \, \frac{ u^{(k)}(x) 
u^{(m)}(x)}{W(k,m)} = -\frac{d}{dx} \ln W(k,m). 
\label{20}
\ee

Notice that the right hand side of (\ref{20}) corresponds to the Crum's
determinant \cite{C55}. Hence, the coefficient of the second term of
$B^{(km)}$ in (\ref{10}) can be constructed either by using (\ref{20}) or
simply by (\ref{17}). Using now (\ref{15}), (\ref{8}) and (\ref{9}), with
$\eta(x)$ as given above, it is easy to show that the function
$u^{(k)}/W(k,m)$ is eigenfunction of $H^{(km)}$ with eigenvalue
$\epsilon^{(m)}$. A similar procedure shows that $ u^{(m)}/W(k,m)$ is also
an eigenfunction of $H^{(km)}$ with eigenvalue $\epsilon^{(k)}$. Taking
this into account we write
\be
\Psi^{(km)}_{\epsilon_m}(x) \propto \frac{u^{(k)}(x)}{W(k,m)}; \quad \quad
\Psi^{(km)}_{\epsilon_k}(x) \propto \frac{u^{(m)}(x)}{W(k,m)},
\label{20a}
\ee
where the subindex $\epsilon_m$ ($\epsilon_k$) indicates the corresponding
eigenvalue $\epsilon^{(m)}$ ($\epsilon^{(k)}$). Therefore, the
eigenfunctions of $H^{(km)}$ are $\{\Psi^{(km)}_{\epsilon_m},
\Psi^{(km)}_{\epsilon_k}\} \cup \{\Psi^{(km)} \propto B^{(km)} \psi \, \,
\vert \, \, B^{(km)} \psi \neq 0 \}$, and the corresponding eigenvalues
are $\{E^{(km)} \} = \{\epsilon^{(m)}, \epsilon^{(k)}, E\}$. Thus,
$H^{(km)}$ has the same spectrum as $H$ plus two new levels at
$\epsilon^{(m)}$ and $\epsilon^{(k)}$, provided that
$\Psi^{(km)}_{\epsilon_m}$ and $\Psi^{(km)}_{\epsilon_k}$ are square
integrable functions. As (\ref{17}) involves two general solutions to the
Riccati equation (\ref{4}), thus $V^{(km)}(x)$ represents a set of {\it
two}-parametric families of potentials almost isospectral to $V(x)$. 

Notice that we have not placed any ordering to the levels $\epsilon^{(m)}$
and $\epsilon^{(k)}$ so that it is possible either that $\epsilon^{(m)}>
\epsilon^{(k)}$ or that $\epsilon^{(m)} < \epsilon^{(k)}$. This is
implicit in the fact that $\eta(x)$ in (\ref{17}) is invariant under the
change $k\leftrightarrow m$. This will be important in the next section,
where we are going to show that the results found by means of the second
order intertwining technique can be found also by the iteration of two
successive first order intertwining transformations. 

\subsection{Iterative factorization}

Let us come back to the results of Section 2.1 and suppose that by means
of the first order intertwining technique we have obtained a Hamiltonian
$H^{(k)}$ from $H$ using a certain factorization energy $\epsilon^{(k)}$.
We look now for a new Hamiltonian $h^{(km)}= - d^2/dx^2 + v^{(km)}(x)$
attainable from $H^{(k)}$ by means of the first order intertwining
method, {\it i.e.}, $h^{(km)}$ and $H^{(k)}$ satisfy the following
relationship
\[
h^{(km)} a^{(km)} = a^{(km)} H^{(k)},
\]
where $a^{(km)}$ is the first order differential operator $ a^{(km)}
\equiv d/dx + \beta^{(km)}(x)$. As usual, we arrive at the standard
equations linking $v^{(km)}(x), \ V^{(k)}(x)$ and $\beta^{(km)}(x)$: 
\be
-(\beta^{(km)})' +(\beta^{(km)})^2 = V^{(k)}(x) -\varepsilon_m = 2
(\beta^{(k)})' + V(x) - \varepsilon_m,
\label{21}
\ee
\be
v^{(km)}(x) = V^{(k)}(x) + 2(\beta^{(km)})'(x) = V(x) + 2 [
\beta^{(km)}(x) +\beta^{(k)}(x)]',
\label{22}
\ee
where $\varepsilon_m$ is a new integration constant (factorization energy) 
and we have used equation (\ref{5}). The intertwined Hamiltonians once
again become factorized
\[
H^{(k)} = {a^{(km)}}^{\dagger}a^{(km)} + \varepsilon_m, \quad \quad
h^{(km)} =a^{(km)}{a^{(km)}}^{\dagger} + \varepsilon_m.
\]
The eigenvalues of $h^{(km)}$ are given by $\{ {\cal E}^{(km)} \} = \{
\varepsilon_m, E^{(k)}\}$, and the eigenfunctions by
$\psi^{(km)}_{\varepsilon_m} \cup \{ \psi^{(km)} = (E^{(k)} -
\varepsilon_m)^{-1/2} \, a^{(km)} \psi^{(k)} \, \, \vert \, \, a^{(km)}
\psi^{(k)} \neq 0 \}$, where the missing state
$\psi^{(km)}_{\varepsilon_m} \propto \exp (\int \beta^{(km)}(x) dx )$
corresponds to the eigenvalue $\varepsilon_m$. 

The key point here becomes to find the general solution to the new
Riccati's differential equation (\ref{21}) for some $\varepsilon_m \in
{\bf R}$. The obvious solution $(\beta^{(km)})'(x) = - (\beta^{(k)})'(x)$,
for which $\varepsilon_m = \epsilon^{(k)}$, corresponds to a Hamiltonian
$h^{(km)}$ equal to $H$ because $v^{(km)}(x)  = V(x)$. Avoiding this
trivial solution we shall consider $\varepsilon_m \neq \epsilon^{(k)}$. 
Non-trivial solutions to (\ref{21}) can be found by inspecting the
following intertwining relationship:
\be
h^{(km)} a^{(km)} a^{(k)} = a^{(km)} H^{(k)} a^{(k)} = a^{(km)} a^{(k)} H.
\label{23}
\ee

Equation (\ref{23}) shows that $h^{(km)}$ and $H$ are related by the
product of two first order intertwining operators, i.e., by the second
order differential operator $C^{(km)} \equiv a^{(km)} a^{(k)}$. An
interesting question immediately arises: is the Hamiltonian $h^{(km)}$ the
same as the $H^{(km)}$ of Section 2.2? If $C^{(km)}= B^{(km)}$ the answer
is affirmative, and in this case the SUSY potential
$\beta^{(k)}(x)$, the {\it iterated} new SUSY potential $\beta^{(km)}(x)$,
and the functions $\gamma(x)$ and $\eta(x)$ are related by the equations
\be
\eta(x) = \beta^{(km)}(x) + \beta^{(k)}(x), \quad \quad \gamma(x) =
\frac{d}{dx}\beta^{(km)}(x) + \beta^{(km)}(x) \beta^{(k)}(x). 
\label{24}
\ee
Comparing (\ref{17}) and (\ref{24}) we get the solutions of (\ref{21}) 
(with $\varepsilon_m = \epsilon^{(m)}$):
\be
\beta^{(km)}(x) = - \beta^{(k)}(x) -\left( \frac{\epsilon^{(m)} -
\epsilon^{(k)}}{\beta^{(m)}(x) - \beta^{(k)}(x)} \right),
\label{25}
\ee
where $\beta^{(k)}(x)$ ($\beta^{(m)}(x)$) is a solution of (\ref{4}) 
associated to the factorization energy $\epsilon^{(k)}$
($\epsilon^{(m)}$). Equation (\ref{25})  is nothing but the theorem of
Fern\'andez {\it et al} \cite{F98}. The eigenfunctions of
$h^{(km)}=H^{(km)}$ are given by the basic set
\[
\Psi^{(km)}_{\epsilon_k}= (\epsilon^{(k)} - \epsilon^{(m)})^{-1/2}
a^{(km)} \psi^{(k)}_{\epsilon_k}, \quad \Psi^{(km)} = (E-
\epsilon^{(m)})^{-1/2} (E - \epsilon^{(k)})^{-1/2} a^{(km)}a^{(k)} \psi,
\]
plus the new missing state 
\[
\Psi^{(km)}_{\epsilon_m} \propto e^{\int \beta^{(km)}(x)dx},
\]
which, by definition, is such that $a^{(km)\dagger}
\Psi_{\epsilon_m}^{(km)} =0$. Hence, by construction $B^{(km)\dagger}
\Psi_{\epsilon_m}^{(km)} = a^{(k)\dagger} a^{(km)\dagger}
\Psi_{\epsilon_m}^{(km)} = 0$. On the other hand, remembering that
$a^{(k)\dagger} \psi_{\epsilon_k}^{(k)}=0$, we get
\[
B^{(km)\dagger} \Psi_{\epsilon_k}^{(km)} = (\epsilon^{(k)} -
\epsilon^{(m)})^{-1/2} a^{(k)\dagger} [ H^{(km)} - \epsilon^{(m)}]
\psi_{\epsilon_k}^{(k)}=0.
\]
Now, by using (\ref{24}), (\ref{8}) and after (\ref{20}), we can write
\[
\Psi_{\epsilon_k}^{(km)}(x) = (\epsilon^{(k)} - \epsilon^{(m)})^{1/2}
\frac{u^{(m)}(x)}{W(k,m)},
\]
which corresponds to the definition of $\Psi_{\epsilon_k}^{(km)}(x)$ given
in (\ref{20a}). A similar procedure but now using (\ref{25}), (\ref{8}) 
and (\ref{20}), allows one to recover the corresponding expression for
$\Psi_{\epsilon_m}^{(km)}(x)$. Finally, the set of eigenvalues of
$h^{(km)} = H^{(km)}$ is given by $\{{\cal E}^{(km)}= E^{(km)} \} = \{
\epsilon^{(m)} , \epsilon^{(k)}, E \}$, just as we have shown in Section
2.2.

It is worth to notice that the operator $B^{(km)}=a^{(km)}a^{(k)}$ does
not factorize the intertwined Hamiltonians $H$ and $H^{(km)}$, but some of
their quadratic forms: 
\bea
B^{(km)\dagger} B^{(km)} &=&
a^{(k)\dagger} a^{(km)\dagger} a^{(km)} a^{(k)} = a^{(k)\dagger} [H^{(k)}-
\epsilon^{(m)}] a^{(k)} = a^{(k)\dagger} a^{(k)} [H- \epsilon^{(m)}]
\nonumber
\\
&=& (H - \epsilon^{(k)})(H - \epsilon^{(m)}) = (H - \epsilon^{(m)})(H -
\epsilon^{(k)}),
\label{26a}
\eea
and
\be
B^{(km)} B^{(km)\dagger} = (H^{(km)} -\epsilon^{(m)} )( H^{(km)} -
\epsilon^{(k)}).
\label{26b}
\ee

Up to now, from the general solution to (\ref{4}) for each factorization
energy $\epsilon^{(k)}$, we have derived a {\it one}-parametric family of
potentials $V^{(k)}(x)$ isospectral to $V(x)$ but by a new level at
$\epsilon^{(k)}$.  From two of these solutions with $\epsilon^{(k)} \neq
\epsilon^{(m)}$ we have derived as well {\it two}-parametric families
isospectral to $V(x)$ but by two new levels at $\epsilon^{(m)}$ and
$\epsilon^{(k)}$. It is possible to iterate further the first order method
by considering now the general solutions to (\ref{4}), associated to three
different factorization energies $\epsilon^{(k)}, \ \epsilon^{(m)}, \
\epsilon^{(l)}$. The key Riccati equation to be solved becomes now: 
\[
-(\beta^{(kml)})'(x) + (\beta^{(kml)})^2 (x) = V^{(km)} (x) -
\epsilon^{(l)}. 
\] 
The algorithm (\ref{25}) allows one to write $\beta^{(kml)}(x)$ in terms
of two different solutions of (\ref{21}),
\[
\beta^{(kml)} (x) = - \beta^{(km)} - \left( \frac{\epsilon^{(l)} -
\epsilon^{(m)}}{\beta^{(kl)}(x) -\beta^{(km)}(x)} \right), 
\] 
which leads to the following 3-parametric family of potentials
$V^{(kml)}(x)$:
\be
V^{(kml)}(x) = V^{(km)}(x) + 2(\beta^{(kml)})'(x) = V(x) + 2 [
\beta^{(kml)}(x) + \beta^{(km)}(x) + \beta^{(k)}(x) ]', 
\label{27}
\ee 
having the same spectrum as $V(x)$ but by three new energy levels at
$\epsilon^{(l)}$, $\epsilon^{(m)}$ and $\epsilon^{(k)}$ \cite{F98}. The
corresponding third order intertwining operator is given by $C^{(kml)} =
a^{(kml)}a^{(km)}a^{(k)}$. This leads to the factorization of some six
order differential operators, related to the two intertwinned Hamiltonians
by means of: 
\[ 
{C^{(kml)}}^{\dagger} C^{(kml)} = (H-\epsilon^{(l)})(H -\epsilon^{(m)})(H-
\epsilon^{(k)}) \] \[ C^{(kml)} {C^{(kml)}}^{\dagger} =
(H^{(kml)}-\epsilon^{(l)})(H^{(kml)} - \epsilon^{(m)})(H^{(kml)} -
\epsilon^{(k)}).
\] 
This process can be continued at will, and the corresponding formulae for
the key functions can be algebraically obtained from the solutions
$\beta^{(k)}(x)$ to (\ref{4}).  It is interesting to see now how the
process does work explicitly for a physically meaningful system.


\section{The hydrogen-like potentials}

Let us consider a single electron in the field produced by a nucleus with
charge $Z{\rm e}$, where $Z$ is the number of protons in the
nucleus\footnote{$Z=1$ describes the Hydrogen atom, while ions or atoms
with only one electron in the outermost shell can be considered as
hydrogen-like systems with $Z>1$.}. The time-independent Schr\"{o}dinger
equation is given by: 
\be
-\nabla^2 \Psi(\vec r) - \frac 2r \Psi(\vec r) = E \Psi (\vec r), 
\label{28}
\ee
where we are using the units of $r_B = \hbar^2/Z e^2 m$, and ${\cal E} =
Z/2 r_B$ for the coordinate and energy respectively.  Due to the spherical
symmetry, the standard choice $\Psi(\vec r)= R(r) Y(\theta, \phi)$
separates (\ref{28})  into a radial and an angular equation.  The
angular solutions are given by the spherical harmonics $Y_l^m(\theta ,
\phi)$ and the energy spectrum can be finally obtained by solving the
radial equation: 
\be
\left[- \frac{d^2}{dr^2} + \frac{l(l+1)}{r^2} - \frac 2r \right] \psi(r)=
E \psi(r),
\label{29}
\ee
where, $l=0,1,2,...$, is the azimutal quantum number and, for the sake of
simplicity, we will work with $\psi(r) \equiv r R(r)$, $0\leq r < +
\infty$. The set of solutions of (\ref{29}) span the whole of $L^2({\bf
R}^+)$, with an inner product defined by $\langle \psi, \psi' \rangle = 4
\pi \int_0^{+\infty}\psi(r)\psi'(r) dr < \infty$.  The differential
operator of the left hand side of (\ref{29}) will be refered as the radial
Hamiltonian, and it will be denoted by $H_l$, where
\be 
H_l = -\frac{d^2}{dr^2} + V_l(x); \quad V_l(x) \equiv \frac{l(l+1)}{r^2} -
\frac 2r.
\label{30}
\ee
The eigenvalues, for a fixed $l$, are given by the well known formula
\be
E_n \equiv E_{lK} = -\frac{1}{(l+K)^2}, \quad K=1,2,3,...
\label{31}
\ee


\subsection{The new one-parametric families}

Let us consider the first order intertwining relationship $H^{(k)}_{l-1}
a^{(k)}_l = a^{(k)}_l H_l$ where, as a result of the further calculations,
we are labeling from the beginning the unknown Hamiltonian $H^{(k)}_{l-1}$
with the subindex $l-1$ because it will arise a centrifugal term for the
new potential with exactly that index. In order to solve the corresponding
Riccati equation (\ref{4}) we shall use its equivalent second order
differential equation (\ref{9}) rewritten as
\be
\left[ -\frac{d^2}{dr^2} + \frac{l(l+1)}{r^2} - \frac{2}{r} \right]
u^{(k)}_l(r)  = \epsilon^{(k)}_l u^{(k)}_l(r). 
\label{32}
\ee
Comparing (\ref{32}) with (\ref{29}) and (\ref{31}), it is natural to
propose the factorization energies as:
\be
\epsilon^{(k)}_l \equiv -\frac{1}{(l+ k)^2}, \quad k \neq K, \quad l>0,
\label{33}
\ee
where, in general, $k$ could be either a discrete as well as a continuous
real number. On the other hand, note that the case $k = K$ reproduces the
well known solutions to the {\it physical} eigenproblem
(\ref{29}-\ref{31}); hence, following Mielnik we have taken $k \neq K$. 
It is clear that, for a fixed value of $l$, we have $\epsilon^{(k)}_l \neq
E_{lK}$, $\forall \, k \neq K$.  

In order to solve (\ref{32}) we make the transformation
\be
u^{(k)}_l(r) = r^{-l} e^{r/(l+k)} \Phi^{(k)}_l(r),
\label{34}
\ee
leading to a confluent hypergeometric equation for $\Phi^{(k)}_l(r)$,
whose general solution for the discrete values $k=0,-1,-2,...,-(l-1)$ is
given by the linear combination of confluent hypergeometric functions
\cite{note4,WG89}: 
\be
\Phi^{(k)}_l(r) = {}_1F_1[k,-2l,-2r/(l+k)] - \nu_{lk} [2r/(l+k)]^{1+2l}
{}_1F_1 [1+k+2l,2+2l,-2r/(l+k)], 
\label{35}
\ee
with
\be
\nu_{lk} \equiv \frac{\Gamma (1 + \vert k \vert )}{\Gamma (2 l + 2)}
\frac{\lambda^{(k)}_l}{(-2l)_{\vert k \vert}}, \quad \quad l>0,
\label{36}
\ee
where $\lambda^{(k)}_l$ is a constant to be determined and
\[
(-2l)_{\vert k \vert} \equiv \frac{\Gamma (-2l + \vert k\vert )}{\Gamma
(-2l)} = (-2l + \vert k \vert -1)(-2l + \vert k \vert -2) \cdot \cdot
\cdot (-2l).
\]
Note that $\Phi^{(k)}_l(r=0)=1$, while its asymptotic behaviour is given
by
\[
\Phi^{(k)}_l(r) \sim \left( \frac{2r}{l-\vert k \vert} \right)^{\vert k
\vert} \frac{1- \lambda^{(k)}_l}{(-2l)_{\vert k \vert}}.
\]
Hence, the solutions (\ref{34}) are divergent at the origin as well as in
the limit $r \rightarrow \infty$, and it is clear that $u^{(k)}_l(r) \not
\in L^{2}({\bf R}^+)$. This is the reason why we consider them as {\it
unphysical} solutions to the Schr\"{o}dinger equation (\ref{29})  with the
{\it atypical} eigenvalues (\ref{33}). Let us remark that the
discreteness of $k$ in (\ref{34}-\ref{36}) leads to the most transparent
case available with the mathematical tools at hand \cite{note4}. The
general solution to the corresponding Riccati equation (\ref{4}) arises
after introducing (\ref{34}) in (\ref{8}), which gives:
\be
\beta^{(k)}_l(r) = \frac{l}{r} - \frac{1}{(l+k)} - \frac{d}{dr} \ln
\Phi^{(k)}_l(r), \quad \quad l>0.
\label{37}
\ee
Using (\ref{5}) and (\ref{37}), we get the first order intertwined
potentials
\be
V^{(k)}_{l-1}(r)= V_{l-1}(r) -2\frac{d^2}{dr^2} \ln \Phi^{(k)}_l(r), \quad
\quad l>0. 
\label{38}
\ee 
The second term of (\ref{38}) is free of singularities if
\be
\lambda^{(k)}_l \in \cases{ (-\infty, 1), & for $\vert k \vert$ even;\cr
\cr
(1, \infty), & for $\vert k \vert$ odd. \cr}
\label{39}
\ee

Hence, the new potentials $V^{(k)}_{l-1}(r)$ have the same singularity at
$r=0$ as $V_{l-1}(r)$, provided that $\lambda^{(k)}_l$ takes only values
in the domain (\ref{39}). Since the second term of $V^{(k)}_{l-1}(r)$
tends to zero when $r \rightarrow 0$ and $r \rightarrow \infty$, we
conclude that $V^{(k)}_{l-1}(r)$ behaves as $V_{l-1}(r)$ at the ends of
$[0, \infty)$. 

Now, the eigenfunctions of $H^{(k)}_{l-1}$ are obtained from the action of
$a^{(k)}_l$ on the eigenfunctions $\{\psi_{nl}(r)\}$ of $H_l$,
$\psi^{(k)}_{n,l-1}= (E_n - \epsilon^{(k)}_l)^{-1/2} \, a^{(k)}_l
\psi_{nl}(r)$, plus the {\it isolated} ground state:
\be
\psi^{(k)}_{l-1, \epsilon_k}= {\cal C}_{lk} \frac{r^l
e^{-r/(l+k)}}{\Phi^{(k)}_l(r)},
\label{40}
\ee
where $\Phi^{(k)}_l(r)$ is given by (\ref{35}), and
\be
{\cal C}_{lk} = \left[ \left(\frac{2}{l -\vert k \vert} \right)^{2l+1}
\left(\frac{1-\lambda^{(k)}_l}{(-2l)_{\vert k\vert}} \right)  \frac{\vert
k \vert !}{(2l)!} \right]^{1/2}, \quad \quad k=0, -1, -2,..., -(l-1).
\label{41}
\ee
It is important to notice that, despites the resemblance between the
missing states (\ref{40}) and the {\it unphysical} functions (\ref{34}),
$\psi^{(k)}_{l-1, \epsilon_k} \sim 1/u^{(k)}_l$, just the first ones
become square integrable functions. The eigenvalues of $H_{l-1}^{(k)}$
(for fixed $l$ and $k$) are:
\be
E_{l-1,K}^{(k)} = -\frac{1}{(l+k)^2}, -\frac{1}{(l+K)^2}, \quad
K=1,2,...
\label{42}
\ee
By comparing with (\ref{31}) it is apparent that our families of
potentials (\ref{38}) have, in general, the same spectrum as the
corresponding spectrum of the hydrogen-like potentials $V_{l-1}(r)$, plus
a new level at $\epsilon_l^{(k)}$. 


\subsection{Particular one-parametric families}

Notice that, for fixed $l$, $k$ can take $l$ different values $k= 0, -1,
-2,..., -(l-1)$. Hence, there are $l$ different factorization energies
$\epsilon^{(k)}_l$ generating $l$ non-equivalent families of solvable
potentials $V_{l-1}^{(k)}(r)$. In particular, let us notice that
$\epsilon^{(0)}_l = -1/l^2\equiv L(l)$, with $\lambda_{l}^{(k)}=0$, lead
to the standard factorization of the hydrogen-like potentials as presented
by Infeld-Hull (see equation 8.1.3, and page 68 of \cite{IH51}).  Thus,
our technique can be considered as a generalization of the Infeld-Hull
factorization where, instead of looking for particular solutions to
(\ref{4}) with $\epsilon^{(0)}_l = L(l)$, we look for the general
solutions with $\epsilon^{(k)}_l = L(l+k)$. In a similar way, our results
can be seen as a generalization of those derived by Fern\'andez, who found
a general solution to (\ref{4}) but just in the case with $k=0$
\cite{F84}.  Hence, for a fixed $l$ and taking $k=0$, the equation
(\ref{42}) leads to $E^{(0)}_{l-1,K} = E_{l-1,K}$, {\it i.e.},
$V^{(0)}_{l-1,}(r)$ is a family of potentials strictly isospectral to
$V_{l-1}(r)$. On the other hand, for $k \neq 0$, we have not only
$\epsilon^{(k)}_{l-1} \neq E_{l-1,K}$, but $\epsilon^{(k)}_{l-1} <
E_{l-1,K=1}= -1/(l+1)^2$. In this case the spectrum of $H^{(k \neq
0)}_{l-1}$ is almost the same as the one of $H_{l-1}$, the difference
resting in the ground state energy level.  It is clear now that the
functions (\ref{38}) represent a set of $l-1$ families of potentials
almost isospectral to $V_{l-1}(r)$, plus a family strictly isospectral to
$V_{l-1}(r)$. 


\subsubsection{The case $k=0$}

The member of the family (\ref{38}), for $k=0$, is given by 
\be
V_{l-1}^{(0)}(r)= V_{l-1}(r) - 2 \frac{d^2}{dr^2} \ln \, \Phi^{(0)}_l(r),
\quad l>0,
\label{43}
\ee 
where 
\be 
\Phi^{(0)}_l(r)= 1 - \left( \frac 2l \right)^{2l+1}
\frac{\lambda^{(0)}_l}{(2l)!} \int_0^r x^{2l}e^{-2x/l} dx.
\label{44}
\ee 
From (\ref{39}) we see that $V_{l-1}^{(0)}(r)$ does not have more
singularities than $V_{l-1}(r)$ if $\lambda^{(0)}_l \in (-\infty, 1)$. In
particular, for $\lambda^{(0)}_l= 0$, we have $\Phi^{(0)}_l(r) = 1$ and
$V_{l-1}^{(0)}(r)= V_{l-1}(r)$, which means that $V_{l-1}(r)$ itself is a
member of the family (\ref{43}). On the other hand, we get also the
standard SUSY potential $W(r) \equiv \beta_l^{(0)}(r) = l/r -1/l$, besides
the corresponding first order operator $a^{(0)}_l = d/dr + l/r - 1/l$,
both of them used to solve the hydrogen-like Hamiltonians
\cite{S40,IH51,W81}.  Let us remark once again that $V_{l-1}^{(0)}(r)$
with $\lambda^{(0)}_l= (2l)! ( l/2)^{2l+1} \gamma^{-1}_l$ leads to the
{\it one}-parametric family of isospectral hydrogen-like potentials
derived by Fern\'andez \cite{F84}. If $l=1$ and $\lambda^{(0)}_1
\rightarrow 1$, the potential (\ref{43}) tends also to the one found by
Abraham and Moses \cite{AM80}. 


\subsubsection{The case $k=-1$}

Let us take now $k=-1$. Thus, the potential (\ref{38}) becomes
\be
V_{l-1}^{(-1)}(r)= V_{l-1}(r) - 2 \frac{d^2}{dr^2} \ln \,
\Phi^{(-1)}_l(r), \quad l>1,
\label{45}
\ee 
where 
\be 
\Phi^{(-1)}_l(r)= \left[1-\frac{r}{l(l-1)}\right] \left\{ 1+
\frac{\lambda^{(-1)}_l}{(2l-1)!} \left(\frac{2}{l-1} \right)^{2l-1}
\int_0^r \frac{x^{2l} e^{-2x/(l-1)}}{[l(l-1) -x]^2} dx \right\}. 
\label{46}
\ee 
In this case $\lambda^{(-1)}_l \in (1, \infty)$, and although $V_{l-1}(r)$
governs the behaviour at the origin and at infinity of
$V_{l-1}^{(-1)}(r)$, it is not a member of the family (\ref{45}). The
corresponding eigenvalues are given by
\be 
E_{l-1,K}^{(-1)} = -\frac{1}{(l-1)^2}, -\frac{1}{(l+1)^2},
-\frac{1}{(l+2)^2}, ..., \quad \quad l>1.  
\label{47}
\ee 
Notice the `jump' between the first two energy levels of $H^{(-1)}_{l-1}$; 
this produces a {\it hole} in the spectrum (\ref{47}) because the level at
$-1/l^2$ is absent.  Comparing (\ref{47}) with (\ref{31}) it turns out
that $V_{l-1}^{(-1)}(r)$ represents a family of potentials almost
isospectral to $V_{l-1}(r)$. As a particular example let us consider
$l=2$: in this case $V^{(-1)}_1(r)$ has the same levels, $-1/n^2$, as
$V_1(r)$ for $n \geq 3$. However, the ground state energy level of
$V^{(-1)}_1(r)$ is at $-1$, which is forbidden for $V_1(r)$ whose ground
state is at $-1/4$. The potentials (\ref{45}-\ref{46}) have been recently
reported by the author (see equations (10) and (22) of \cite{R98}) and
they are plotted in Figure 1 for $l=2$ and different values of
$\lambda_2^{(-1)}$. 


\subsubsection{The case $\vert k \vert > 1$}

For $\vert k \vert >1$, there is a gap between the two first energy levels
of $H_{l-1}^{(k)}$, which grows up as $\vert k \vert$ increases. For
instance, when $\vert k \vert =2$, and $l=3$, the levels labeled by $n=2$
and $n=3$ are not present in the spectrum of $H_2^{(-2)}$. A similar
result, but with a different method and Hamiltonian, has been derived by
Samsonov \cite{AB85}.

Recently, the SUSY potential $\omega(x) = a/\gamma - \gamma/x$, with $a>0$
and $\gamma >0$, has been used to construct a `conditionally exactly
solvable (CES) potential' $V_+(x)$ for the Hydrogen atom problem
\cite{note5,JR98}. In such a paper $a$ is a coupling constant while
$\gamma$ is a parameter taking some strange values as $\gamma = 2.8$ (see
Fig. 7 in \cite{JR98}). Although there is a close relationship between
that potential $V_+(x)$ and our first order intertwined potential
$V_{l-1}^{(k)}(r)$, it is important some caution: 
\begin{enumerate}
\item[]
{\it i}) If $\gamma >0$ is not an integer, the SUSY potential $\omega(x)$
of \cite{JR98} is not related to the physically relevant SUSY potential
$W(r)$ for the Hydrogen atom, and in this case $V_+(x)$ is the `CES
potential' for some {\it mathematical problem}.  If the so called CES
potentials derived in \cite{JR98} are to be related with a ``physical
potential", the condition $\gamma >0$ is not sufficient. 

\item[]
{\it ii}) If $\gamma$ is an integer greater than zero, the SUSY potential
$\omega(x)$ can be related to the Hydrogen atom problem. In this case the
function (5.14) of \cite{JR98} has a term ${}_1F_1(-\gamma -a/\rho, -2
\gamma, 2 \rho x)$, which is meaningless at $2 \rho x =0$ unless $\gamma +
a/\rho$ is an integer greater than or equal to zero \cite{WG89}.  Indeed,
since $\gamma$ is a positive integer, the fraction $a/\rho$ is an integer
such that $\gamma \geq -a/\rho$.  Hence, the condition $a>0$ is neither
sufficient to ensure the generality of the potential $V_+(x)$.
\end{enumerate}

Taking this into account, a general `CES potential' $V_+(x)$ really
related to the Hydrogen atom problem would take for instance $-a/\rho =
\gamma + k$, where $k$ is an integer such that $k \leq 0$. In particular,
for $\gamma = l$, and $x=r/a$, the CES potential $V_+(x)$ derived in
\cite{JR98} would be equal to our first order intertwined potential
$V_{l-1}^{(k)}(r)$, provided that $\nu_{lk}=\beta/\alpha$, where $\alpha$
and $\beta$ are constants.

We are going to derive next some {\it two}-parametric families by means of
the technique discussed in sections 2.2 and 2.3 for the particular case we
are dealing with.


\subsection{ The new two-parametric families}

In Section 2.2 we have seen that the second order intertwining
relationship $H^{(km)}_{l-2} B^{(km)}_{l-2} = B^{(km)}_{l-2} H_l$ provides
new solvable potentials if we are able to find some solutions $\eta(x)$ to
equation (\ref{12}). Then we also have shown that the general form of that
solutions is given by: 
\be
\eta(r) = (\epsilon^{(k)}_l- \epsilon^{(m)}_l) \left\{ \frac{d}{dr} \ln
\,\left( \frac{u^{(k)}_l(r)}{u^{(m)}_l(r)} \right) \right\}^{-1}, \quad
\quad \epsilon^{(k)}_l \neq \epsilon^{(m)}_l.
\label{48}
\ee
For our particular system we take the functions $u^{(k)}_l(r)$ as given by
(\ref{34}). The behaviour of our $\eta(r)$ at the ends of $[0,\infty)$ is
given by
\[
\eta(r)_{r \rightarrow \infty} \sim - \frac{2l+k+m}{(l+k) (l+m)}, \quad
\quad \eta(r)_{r \rightarrow 0} \sim -\frac{(1-2l)}{r}.
\]
Hence, it is natural to write the new potential $V^{(km)}_{l-2}(r)$ as
(see equation (\ref{12})):
\be
V_{l-2}^{(km)}(r) = V_{l-2}(r) + 2 \alpha'(r),
\label{49}
\ee
where $\alpha(r) \equiv \eta(r) + (1-2l)/{r}$ is an appropriate function
making evident the limit $V^{(km)}_{l-2}(r) \rightarrow V_{l-2}(r)$ when
$r \rightarrow + \infty$, or $r \rightarrow 0$. The {\it two}-parametric
domain of $\lambda^{(k)}_l$ and $\lambda^{(m)}_l$, for which $\alpha'(r)$
is free of singularities, is determined by the parities of $k$ and $m$. In
order to make transparent the choice of that domain we take the following
convention: $n$ is the index labelling the factorization energy
$\epsilon^{(n)}_l$ defined by $\epsilon^{(n)}_l \equiv {\rm
max}(\epsilon^{(k)}_l, \epsilon^{(m)}_l)$, and $s$ is such that
$\epsilon^{(s)}_l \equiv {\rm min}(\epsilon^{(k)}_l, \epsilon^{(m)}_l)$.
Hence, we have:
\be
\cases{
\lambda^{(n)}_l \in (-\infty, 1), & $\lambda^{(s)}_l \in (-\infty, 1)$ ; 
\, $\vert n \vert$ even, $\vert s \vert$ odd \cr
\cr 
\lambda^{(n)}_l \in (1,\infty), & $\lambda^{(s)}_l \in (1,\infty)$ ; \, \,
\, $\vert n \vert$ odd, $\vert s \vert$ even. \cr}
\label{50}
\ee
\bigskip
\be
\cases{
\lambda^{(n)}_l \in (-\infty, 1), & $\lambda^{(s)}_l \in (1,\infty) $ ; \,
\, \,  $\vert n \vert$ even, $\vert s \vert$ even \cr
\cr
\lambda^{(n)}_l \in (1,\infty), & $\lambda^{(s)}_l \in (-\infty, 1)$ ; \,
\, $\vert n \vert$ odd,  $\vert s \vert$ odd \cr}
\label{51}
\ee
The eigenfunctions of $H^{(km)}_{l-2}$ are given by
\be
\Psi_{n, l-2}^{(km)} (r) = \frac{B^{(km)}_{l-2} \psi_{nl}(r)}
{\sqrt{(E_{n}-\epsilon^{(m)}_l) (E_{n}-\epsilon^{(k)}_l)}},
\label{52}
\ee
and
\be
\Psi_{\epsilon_{k},l-2}^{(km)} (r) = {\cal C}_{lk} \sqrt{\epsilon^{(k)}_l-
\epsilon^{(m)}_l} \, \, \frac{u^{(m)}_l(r)}{W(k,m)}, \quad \quad
\Psi_{l-2, \epsilon_{m}}^{(km)} (r) = {\cal C}_{lm}
\sqrt{\epsilon^{(m)}_l- \epsilon^{(k)}_l} \, \,
\frac{u^{(k)}_l(r)}{W(m,k)},
\label{53}
\ee
where ${\cal C}_{lk}$ is given by (\ref{41}). The corresponding
eigenvalues are: 
\be 
E_{l-2,K}^{(km)} = \left\{ -\frac{1}{(l+m)^2}, -\frac{1}{(l+k)^2} \right\}
\cup \left\{-\frac{1}{(l+K)^2} \right\}, \quad \quad K>0.
\label{54}
\ee

Let us remark that the only restriction for the factorization energies,
$\epsilon_l^{(k)} \neq \epsilon_l^{(m)}$, leads to the fact that $\eta
(r)$ and $V_{l-2}^{(km)}(r)$ are symmetric under the change $k \rightarrow
m$ and {\it vice versa}, $V_{l-2}^{(km)}(r) = V_{l-2}^{(mk)}(r)$. However,
this symmetry is broken for the intermediate {\it one-}parametric
potentials arising when $V_{l-2}^{(km)}(r)$ is obtained after two iterated
first order intertwining procedures. In order to see that, let us suppose
that $\epsilon_l^{(k)} > \epsilon_l^{(m)}$. We have two different ways to
get the potentials (\ref{49}): we can add first the level
$\epsilon_l^{(k)}$, and then the level $\epsilon_l^{(m)}$; the other
option is to perform the two operations in the opposite order. 

\begin{itemize}
\item[]
({\bf I}) In the first case, after the first step we shall get the
Hamiltonian $H_{l-1}^{(k)}$, with a spectrum equal to the spectrum of
$H_l$ plus a new level at $\epsilon_l^{(k)} \leq E_{l,K=1}$, where
$E_{l,K=1} = -1/(l+1)^2$ is the ground state energy level of $H_l$. After
the second step, the final Hamiltonian $H_{l-2}^{(km)}$ has the same
spectrum as $H_{l-1}^{(k)}$ plus a new level at $\epsilon_l^{(m)} <
\epsilon_l^{(k)}$.  In both of these procedures the factorization energy
(the ground state energy of the new Hamiltonian) is less than the ground
state energy of the initial Hamiltonian, just as it happens in the SUSY
approach. 
\item[]
({\bf II}) Taking now the two operations in the opposite order, the first
intertwined Hamiltonian $H_{l-1}^{(m)}$ has a spectrum equal to the one of
$H_l$ plus a new level at $\epsilon_l^{(m)} \leq E_{l,K=1}$, in agreement
once again with the SUSY approach. However, the next intertwined
Hamiltonian $H_{l-2}^{(mk)}$ is derived by adding a new level at
$\epsilon_l^{(k)} > \epsilon_l^{(m)}$, {\it i.e.}, {\it above} the ground
state energy level of $H_{l-1}^{(m)}$, which is against the SUSY {\it
doctrine} but it is valid in our treatment.
\end{itemize}

A final point is that the equality of the two Hamiltonians
$H_{l-2}^{(km)}$ and $H_{l-2}^{(mk)}$, gotten of two iterations of the
first order transformations, leads to two different factorizations of the
global second order intertwining operator
\[
B_{l-2}^{(km)} = a_{l-1}^{(km)} a_l^{(k)} = a_{l-1}^{(mk)} a_l^{(m)} =
B_{l-2}^{(mk)}
\]


\subsection{Particular two-parametric families}

The simplest {\it two-}parametric family of potentials (\ref{49}) 
corresponds to $k=-1$ and $m=0$. In this case, by means of (\ref{48}) we
get the solution to (\ref{11}): 
\be
\eta(r) = \frac{1-2l}{l^2(l-1)^2} \left\{ \frac{d}{dr} \ln \,\left(
\frac{e^{r/l(l-1)} 
\Phi^{(-1)}_l(r)}{\Phi^{(0)}_l(r)} \right) \right\}^{-1},
\quad \quad l>1,
\label{55}
\ee
which has been recently reported in \cite{R98}, where $\Phi^{(0)}_l(r)$,
and $\Phi^{(-1)}_l(r)$ are given by (\ref{44}) and (\ref{46}),
respectively. The appropriate {\it two-}parametric domain (see (\ref{50}) 
and (\ref{51}))  is given by $\lambda^{(-1)}_l, \lambda^{(0)}_l \in
(-\infty, 1)$. From (\ref{54}) and (\ref{31})  it is clear that
$V_{l-2}^{(-1,0)}(r)$ is strictly isospectral to $V_{l-2}(r)$. Notice that
$ \lambda^{(-1)}_l= \lambda^{(0)}_l =0$ leads to $V_{l-2}^{(-1,0)}(r) =
V_{l-2}(r)$. Moreover, for $\lambda^{(0)}_l=0$ and $\lambda^{(-1)}_l =
[(l-1)/2]^{2l-1} (2l-2)!/\gamma_{l-1}$, the potentials (\ref{49}) become
the family (\ref{43}) derived by Fern\'andez, provided $l$ is changed by
$l+1$. We make concrete now the discussion of the cases ({\bf I}) and
({\bf II}) of Section 3.3. 
\begin{itemize} 
\item[] 
({\bf I}) Departing from the Hamiltonian $H_l$, by means of a first order
intertwining operator we get the Hamiltonian $H_{l-1}^{(0)}$, whose
potential is given by (\ref{43}), where we have added the new level
$\epsilon^{(0)}_l = -1/l^2$ below the ground state energy level of $H_l$,
$E_{l, K=1}=-1/(l+1)^2$. The iteration of this procedure leads to
$H_{l-2}^{(0,-1)}$, whose potential is given by (\ref{49}) and (\ref{55}). 
In this second step we have added the new level $\epsilon_l^{(-1)} =
-1/(l-1)^2$ below the previous ground state energy level $\epsilon_l^{(0)}
= -1/l^2$, which agrees with the usual SUSY asumption.
\item[]
({\bf II}) We depart now from $H_l$ to get $H_{l-1}^{(-1)}$, whose
potential is given by (\ref{45}). In this step, we have added the new
level $\epsilon^{(-1)}_l = -1/(l-1)^2$ below the ground state energy level
of $H_l$ but we have left a {\it hole} to be filled during the second step
(see Section 3.2.2). The next step provides now $H_{l-2}^{(-1,0)}$, whose
potential is given again by (\ref{49}) and (\ref{55}). Notice that
$\beta_{l - 1}^{(-1,0)}(r)  = \beta_{l-1}^{(0,-1)}(r)+
\beta_l^{(0)}(r)-\beta_l^{(-1)}(r)$. The new energy level
$\epsilon_l^{(0)} = -1/l^2$ is now added {\it above} the ground state
energy level $\epsilon_l^{(-1)} = -1/(l-1)^2$ of $H_{l-1}^{(-1)}$ but
below the ground state energy level $-1/(l+1)^2$ of $H_l$. This last
procedure fills the {\it hole} generated during the first step of the
whole procedure which is not typical in the usual SUSY asumption. 
\end{itemize}

A different choice of $k$ and $m$ produces families whose spectra are
almost equal to the corresponding hydrogen-like spectra, with the two
first energy levels different from the levels of the corresponding
hydrogen-like potential. For instance, if $k=-3$, $m=0$, and $l=4$, the
potential $V_2^{(-3,0)}(r)$ has a spectrum of the form $-1/n^2$ but
without the levels at $n=2$ and $n=3$, similarly as it happens in the
first order case mentioned in Section 3.2.3. This is illustrated in Figure
2. A more exotic case is obtained if $k=-4$, $m=-1$, and $l=5$, where the
levels with $n=2$, $n=3$, and $n=5$ are not present for the potential
$V_3^{(-4,-1)}(r)$.

\section{Concluding remarks}

The $nth$-order intertwining technique developed in this paper is intended
to derive $n$-parametric families of potentials which can be isospectral,
or almost isospectral to some well known potential. In particular, we have
successfully derived new potentials from the hydrogen-like atom for the
one parametric and two parametric cases. We have shown also that the
derivation of {\it multi}-parametric families of almost isospectral
potentials becomes mathematically easy by the iteration of the first order
intertwining technique. Although the only trouble in our approach could be
to find the solutions to a set of $n$ Riccati type equations,
characterized by $n$ different factorization energies, we have shown that
the eigenvalues problem for the initial potential allows one to get that
solutions by means of {\it unphysical} eigenfunctions. Our {\it
one}-parametric results for the hydrogen-like atom show how the technique
effectively works, moreover, our resulting families of potentials are more
general than other results previously derived \cite{S40,IH51,F84,W81}. On
the other hand, our {\it two}-parametric potentials represent a further
generalization of the {\it one}-parametric cases presented in this paper.
Finally, we have shown that the construction of a SUSY partner Hamiltonian
using a factorization energy $\epsilon$ less than the ground-state energy
of the departure Hamiltonian is not a general rule in SUSY QM. In
particular, the generation of {\it holes} in the spectra of the
$n$-parametric families of almost isospectral hydrogen-like potentials
becomes the key-stone in the construction of the {\it atypical} $n$-SUSY
partners of the Hydrogen potential derived in this paper. These techniques
can be applied to other physically interesting situations, e.g., for
systems with continuous spectra as the free particle. A detailed study in
this direction will be given elsewhere \cite{M98}. 

\section*{Acknowledgements}

This work has been supported by CONACyT, Mexico ({\it Programa de
Estancias Posdoctorales en Instituciones del extranjero 1997-1998}). The
author would like to thank to Dr. D. J.  Fern\'andez for fruitful
discussions and suggestions. This work has been benefited from the
comments of Dr. L.~M.~Nieto and has been partially supported by Junta de
Castilla y Le\'on, Spain (project C02/97).  The kind hospitality at
Departamento de F\'{\i}sica Te\'orica and the suggestions of the referees
are also aknowledged.

\vfill\eject

\newpage

\begin{figure}
\centerline{
\epsfbox{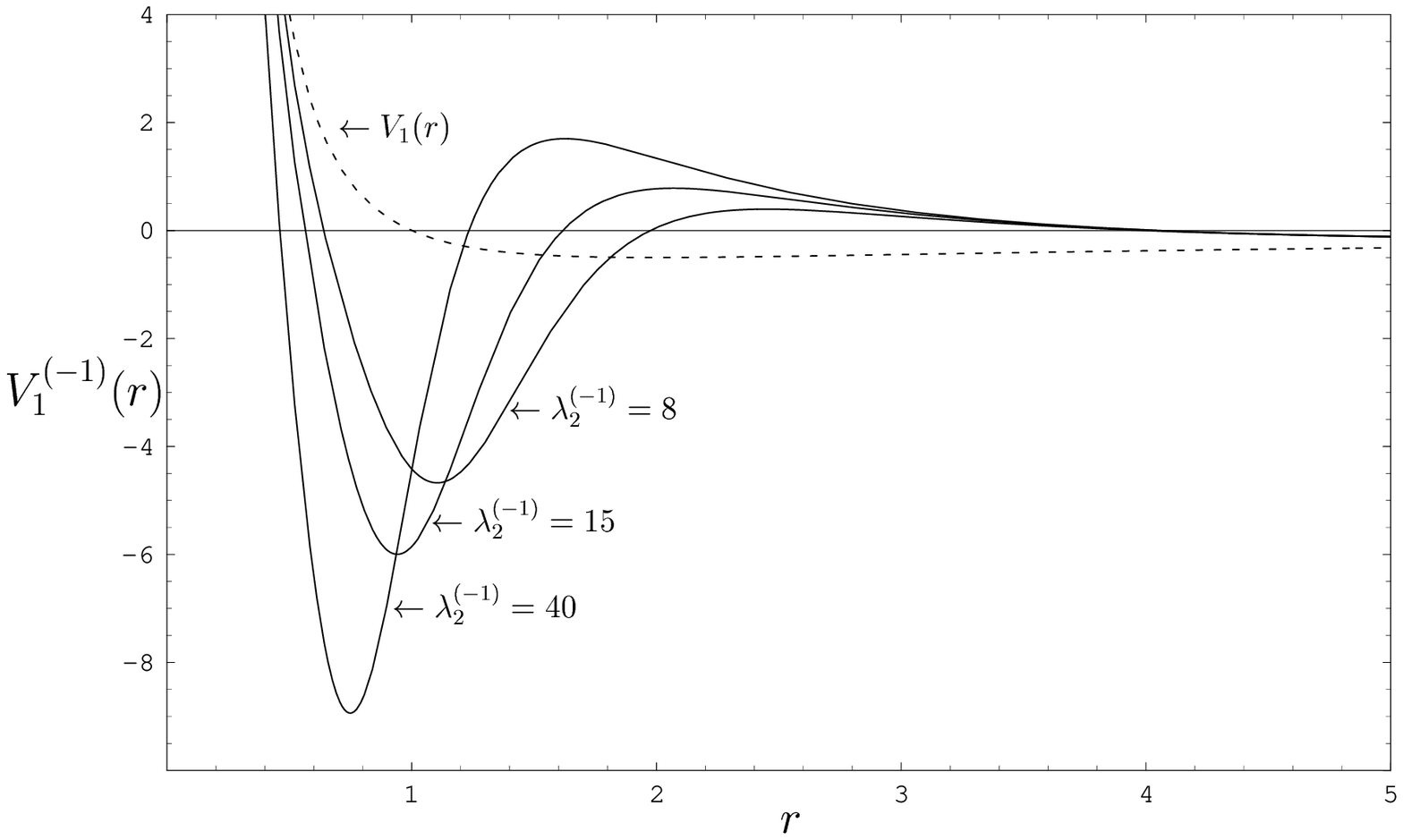}
}
\caption{
 Some members of the {\it one}-parametric family of potentials (\ref{45}) 
for $l=2$ and the indicated values of $\lambda_2^{(-1)}$. The broken curve
represents the hydrogen-like potential $V_{l-1}(r)$. Notice the arising of
global minimum for $V_1^{(-1)}(r)$, which produces a well which can be
modified by changing the values of the parameter $\lambda_2^{(-1)}$. }

\end{figure}  

\begin{figure}
\centerline{
\epsfbox{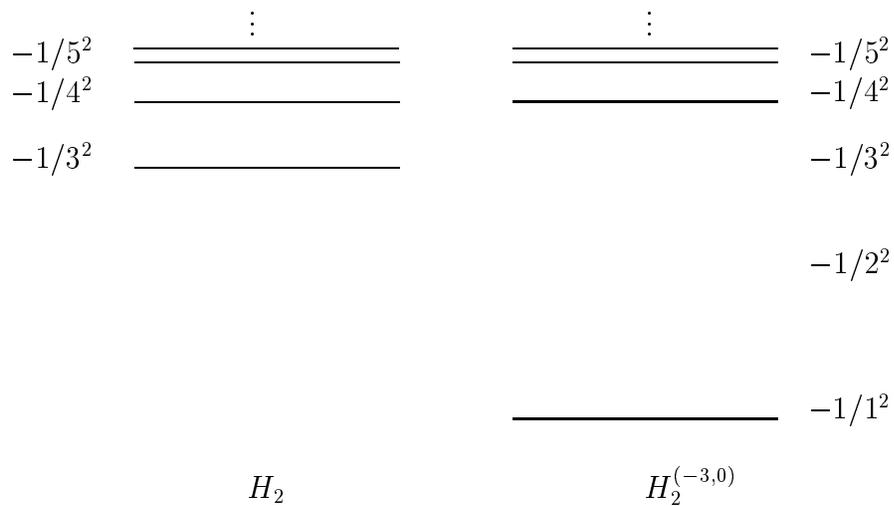}
}
\caption{
Diagram showing the first energy levels of $H_2$ and $H_2^{(-3,0)}$. 
Notice the gap between the ground state energy level
$\epsilon_4^{(-3)}=-1$ and the first excited state
$\epsilon_4^{(0)}=-1/4^2$ of the {\it two}-parametric family of
Hamiltonians $H_2^{(-3,0)}$.  }
\end{figure}


\begin{thebibliography}{99}
\bibitem{note4}
The case when $k$ is different from a negative integer or zero will be
discussed elsewhere.
\bibitem{note5}
For the sake of clarity in the further discussions we have not used the
original notation for $\omega(x)$ as it has been given in \cite{JR98},
where it is $W(x)$.
\end{thebibliography}
\end{document}